# Fractional form of the 0.7(2e2/h) feature

**N T Bagraev**[1]**, N G Galkin**[1]**, W Gehlhoff**[2]**, L E Klyachkin**[1]**, AM Malyarenko**[1] **and I A Shelykh**[3]

[1] Ioffe Physico-Technical Institute RAS, 194021, St.Petersburg, Russia
[2] Technische Universitaet Berlin, D-10623, Berlin, Germany
[3] School of Physics & Astronomy, University of Southampton, SO17 1BJ Southampton, UK

E-mail: impurity.dipole@mail.ioffe.ru

**Abstract**. We present the first findings of the spin transistor effect in the Rashba gate-controlled ring embedded in the p-type self-assembled silicon quantum well that is prepared on the n-type Si (100) surface. The coherence and phase sensitivity of the spin-dependent transport of holes are studied by varying the value of the external magnetic field and the top gate voltage that are applied perpendicularly to the plane of the double-slit ring and revealed by the Aharonov-Bohm (AB) and Aharonov-Casher (AC) conductance oscillations, respectively. Firstly, the amplitude and phase sensitivity of the *0.7·(2e$^2$/h)* feature of the hole quantum conductance staircase revealed by the quantum point contact inserted in the one of the arms of the double-slit ring are found to result from the interplay of the spontaneous spin polarization and the Rashba spin-orbit interaction (SOI). Secondly, the values of the AC conductance oscillations caused by the Rashba SOI are found to take the fractional form with both the plateaus and steps as a function of the top gate voltage.

The spin-correlated transport in low-dimensional systems was in focus of both theoretical and experimental activity in the last decade [1,2]. The studies of the Rashba spin-orbit interaction (SOI) that results from the structure inversion asymmetry in mesoscopic nanostructures have specifically attracted much of the efforts [1,3,4]. The variations in the Rashba SOI value appeared to give rise to the developments of spintronic devices based on the spin-interference phenomena that are able to demonstrate the characteristics of the spin field-effect transistor (FET) even without ferromagnetic electrodes and external magnetic field [4,5]. For instance, the spin-interference device shown schematically in figure 1a represents the Aharonov-Bohm (AB) ring covered by the top gate electrode, which in addition to the geometrical Berry phase provides the phase shift between the transmission amplitudes for the particles moving in the clockwise and anticlockwise direction [4]. This transmission phase shift (TPS) seems to be revealed by the Aharonov-Casher (AC) conductance oscillations measured by varying the top gate voltage applied to the two-terminal device, with the only drain and source constrictions [5]. However, the variations in the density of carriers that accompany the application of the top gate voltage are also able to result in the conductance oscillations. Therefore the three-terminal device with the quantum dot (QD) [6] or the quantum point contact (QPC) [7] inserted in one of the ring's arms using the split-gate technique [8] could be more appropriate to divide the relative contribution of the AC effect in the conductance oscillations (see figure 1a).
Since the AB ring's conductance has to be oscillated with the periodicity of a flux quantum h/e when a variable magnetic field threads its inner core, these AB oscillations have been shown to be persisted, if the transport through the QD [6] or QPC [7] inserted is coherent. The TPS caused by the QD or QPC has been found to be equal to π in the absence of the spin polarization of carriers [6,7], whereas the TPS value π/2 appeared to be very surprisingly in the range of the *0.7·(2e$^2$/h)* feature of the quantum conductance staircase revealed by the QPC inserted thereby verifying the spin polarization in the AB ring [7]. This TPS is of interest to cause no any changes in the amplitude of the *0.7·(2e$^2$/h)* feature, thus remaining unsolved the question on the relative contribution of the spontaneous spin polarization and the Rashba SOI to its creation. Nevertheless, the high sensitivity of the *0.7·(2e$^2$/h)* feature to the variations of the spin polarization of carriers allowed the three-terminal device with the QPC inserted to study the AC conductance oscillations.

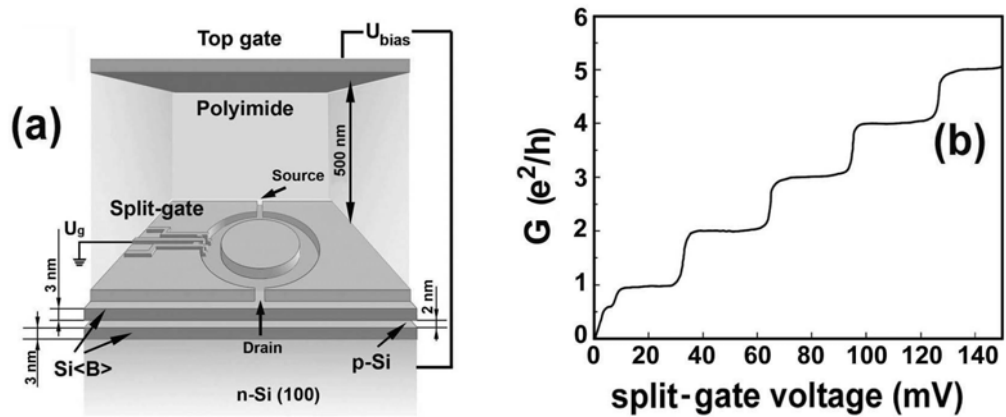

**Figure 1.** (a) Schematic diagram of the device that demonstrates a perspective view of the p-type silicon quantum well confined by the δ - barriers heavily doped with boron on the n-type Si (100) surface, the top gate that is able to control the sheet density of holes and the depletion regions created by the split-gate method, which indicate the double-slit ring with QPC inserted in one of its arms. (b) The quantum conductance staircase as a function of the split-gate voltage, which is revealed by heavy holes tunneling through the QPC inserted in the arm of the Si-based double-slit ring in zero magnetic field and under zero bias voltage controlled by the top gate (see figure 1a).

Here we report the measurements of the amplitude and the TPS of the *0.7·(2e$^2$/h)* feature shown in figure 1b by tuning the Rashba SOI in the three-terminal silicon one-dimensional ring (figure 1a). The device is based on the ultra-narrow, 2 nm, self-assembled silicon quantum well (Si-QW) of the p-type that is prepared on the n-type Si (100) surface by the short-time diffusion of boron from the gas phase. The one-dimensional ring embedded electrostatically in the Si-QW, R=2500 nm, contains the source and the drain constrictions that represent QPCs as well as the QPC inserted in its arm by the split-gate method. The parameters of the high mobility Si-QW were defined by the Hall measurements as well as by the SIMS, STM cyclotron resonance and EPR methods. The initial value of the sheet density of 2D holes, $4·10^{13}$ m$^{-2}$, was changed controllably over one order of magnitude, between $5·10^{12}$ m$^{-2}$ and $9·10^{13}$ m$^{-2}$, by biasing the top gate above a layer of insulator, which fulfils the application of the p$^+$-n bias junction. The variations in the mobility measured at 3.8 K that corresponded to this range of the $p_{2D}$ values appeared to occur between 80 and 420 m$^2$/vs (see figures 2a and 2b). Thus the value of the mobility was high even at low density of the 2D holes. Besides, the high value of mobility showed a decrease no more than two times in the range of temperatures from 3.8 K to 77 K that seems to be caused by the ferroelectric properties for the δ – barriers heavily doped with boron [9], which confine the Si-QW (figure 1a). These characteristics of the 2D gas of holes allowed the studies of the quantum conductance staircase revealed by the heavy holes at 77 K (figure 1b). The number of the highest occupied mode of the QPC inserted in the arm of the AB double-slit ring was controlled by varying the split-gate voltage, whereas the Rashba SOI was tuned by biasing the top gate. The experiments are provided by the effective length of the QPC inserted, 0.2 μm, and the cross section of the 1D channel, 2 nm × 2 nm, which is determined by the width of the Si-QW and the lateral confinement due to ferroelectric properties for the δ – barriers. We focus then on the effects of the bias voltage controlled by the top gate and the external magnetic field on the amplitude and the TPS of the $0.7 \cdot (2e^2/h)$ feature at the fixed value of the split-gate voltage, 5.7 mV, in its range (see figure 1b).
The presence of the Rashba SOI in the device studied is revealed by the measurements of the magnetoconductance (figures 3a and 3b). By varying the top gate voltage at the fixed value of the split-gate voltage, 5.7 mV, the transition from the positive magnetoresistance to the negative magnetoresistance is observed thereby verifying a crossover from the weak antilocalization to the weak localization following the changes of the concentration of the 2D holes. The dependencies shown in figures 3a and 3b are in a good agreement with the theoretical predictions [10] and correlate with the experimental electron magnetoconductance data in a high mobility $In_xGa_{1-x}As/InP$ quantum well [11]. Thus, the SOI effects observed in the studies of the magnetoconductance are evidence of the constructive and destructive backscattering associated with QPCs by varying the top gate voltage thereby allowing the findings of the AC conductance oscillations in the range of the external magnetic fields outside the region of the weak antilocalization.

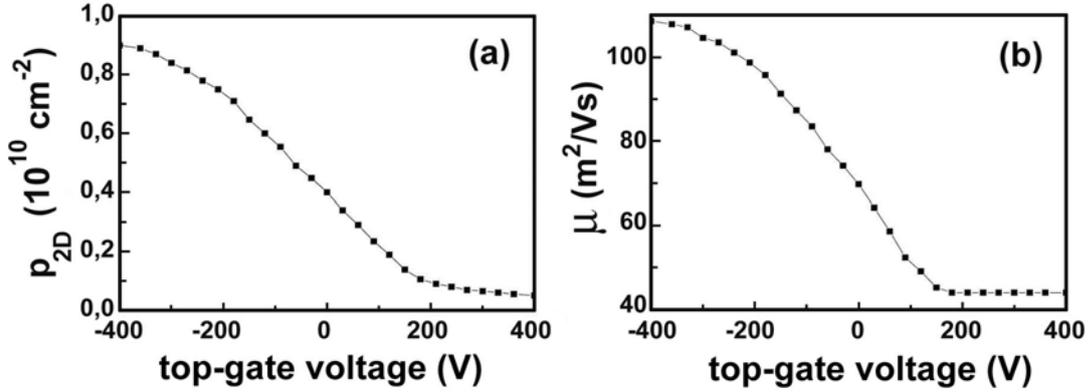

**Figure 2.** Density (a) and mobility (b) revealed by the two-dimensional gas of holes as a function of the top gate voltage which were extracted from longitudinal and Hall voltage measurements of the p-type Si-QW confined by the δ-barriers on the n-type Si (100) surface.

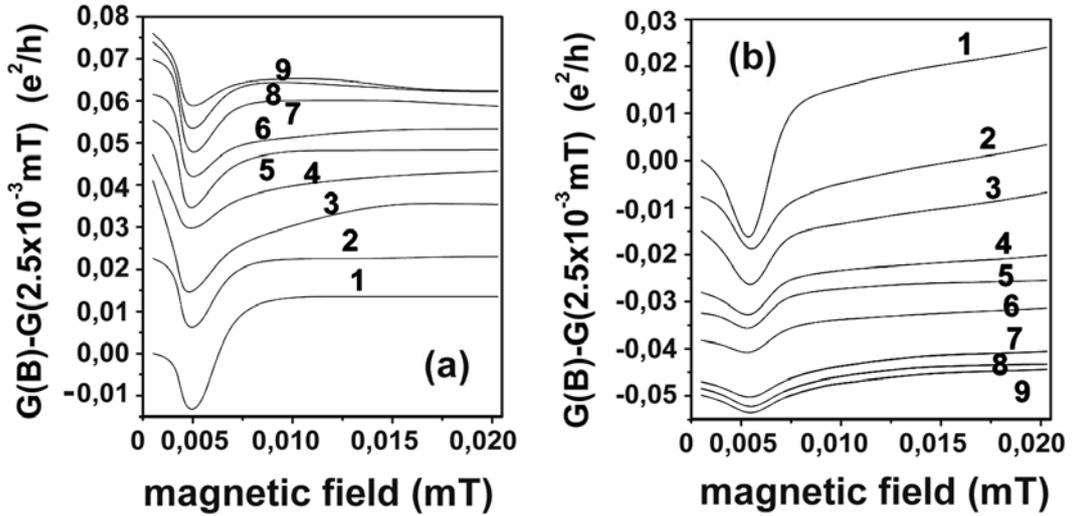

**Figure 3.** Experimental magnetoconductance, $G(B)-G(2.5 \cdot 10^{-3} mT)$, for different negative (a) and positive (b) $U_{bias}$ determined by the top gate voltage. B=$2.5 \cdot 10^{-3}$mT is the residual magnetic field obtained after screening the Earth magnetic field. A crossover from the weak localization to the weak antilocalization is revealed by varying the top gate voltage from negative to positive $U_{bias}$.
(a) $U_{bias}$, mV: 1 – 40, 2 – 140, 3 – 230, 4 – 260, 5 – 300, 6 – 330, 7 – 370, 8 – 390, 9 – 400.
(b) $U_{bias}$, mV: 1 – 30, 2 – 90, 3 – 140, 4 – 230, 5 – 260, 6 – 300, 7 – 360, 8 – 370, 9 – 380.

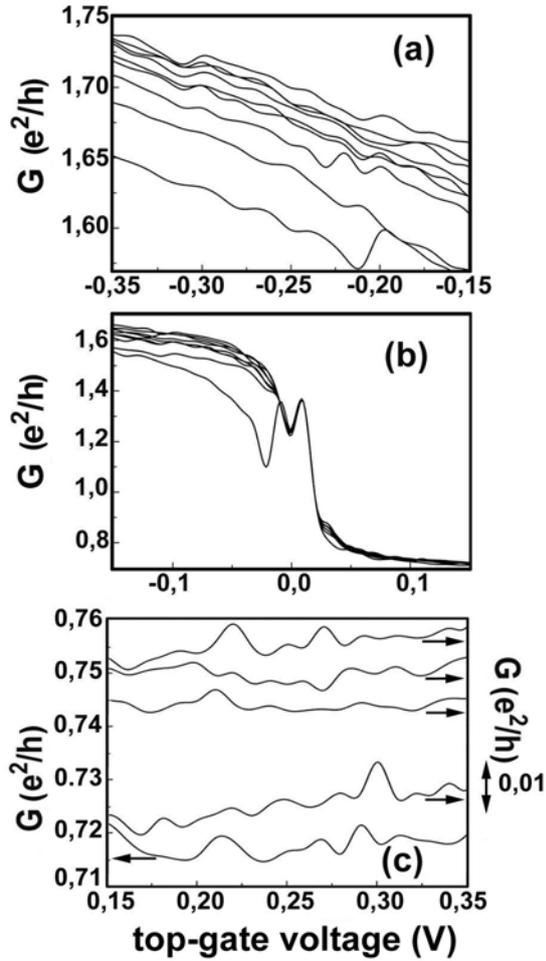

**Figure 4.** The AC conductance oscillations (a, c) that attend the changes in the amplitude of the *"0.7·(2e²/h)"* feature (b) in the double-slit ring with QPC inserted in one of its arms. The external magnetic field value was changed from 0.05 mT (bottom) to 0.5 mT (top) in 0.05 mT step.

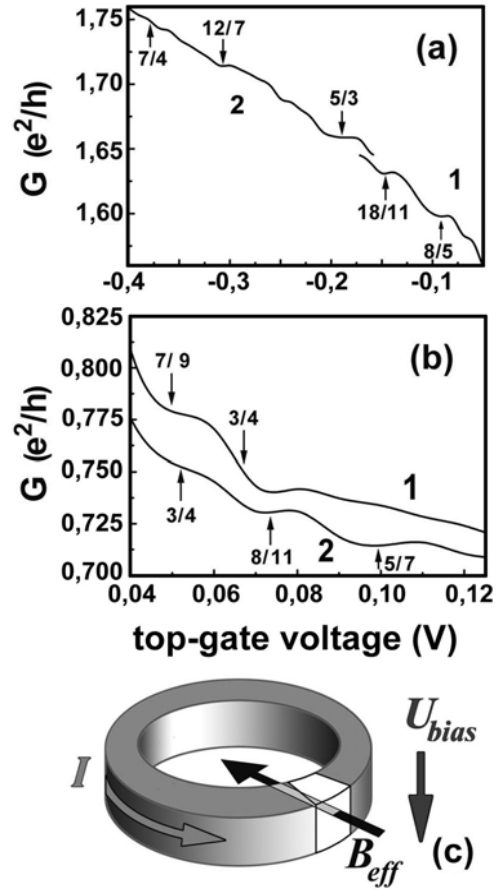

**Figure 5.** (a), (b) The fractional values of the conductance that are revealed by varying the value of the Rashba SOI controlled by the top gate voltage.
(a) 1-B=0.157 mT; 2-B=0.253 mT;
(b) 1-B=0.0025 mT; 2-B=0.0080 mT.
(c) The scheme of the AB ring with QPC. The top gate voltage, $U_{bias}$, controls the density of holes and the effective magnetic field, $B_{eff}$, induced by the Rashba SOI.

Figures 4a and 4b show the changes in the amplitude of the *0.7·(2e²/h)* feature found by biasing the top gate that appeared to be followed by increasing and decreasing the $p_{2D}$ value from the initial value, $4·10^{13}$ m⁻², with applying respectively the reverse and forward bias (see figure 2a). At low density of the 2D holes, the amplitude of the *0.7·(2e²/h)* feature reduces under the forward bias up to the *0.7·(e²/h)* value crossing the point of *0.5·(2e²/h)* that indicates the spin degeneracy lifting for the first step of the quantum conductance staircase [8]. These data seem to be evidence of the spontaneous spin polarization of heavy holes in the 1D channel that is due to the efficient quenching of the kinetic energy by the exchange energy of carriers [9]. The amplitude of the *0.7·(2e²/h)* feature fixed by the split-gate voltage is also found to exhibit the AC conductance oscillations in the absence of changes in the $p_{2D}$ value that are revealed specifically by varying the forward bias voltage applied to the top gate, (see figure 4c). These phase variations of the AC oscillations were shown to be caused by the elastic scattering of the heavy holes on the QPCs inside the double-slit ring [12,13]. The phase shifts calculated in frameworks of this model appeared to be dependent on the Rashba parameter, $\alpha$, determined by the effective magnetic field, which is created by the Rashba SOI,

$$\mathbf{B}_{eff} = \frac{\alpha}{g_B \mu_B} [\mathbf{k} \times \mathbf{e}_z] \qquad (1)$$

and consequently on the modulation of the conductance [14,15], which is found to be in a good agreement with the AC conductance oscillations shown in figure 4c.

The variations of the top gate voltage are very surprisingly to give rise to the fractional form of the conductance (see figures 5a and 5b). The plateaus and steps observed are of interest to bring into correlation with the odd and even fractions that seem to be caused by the geometry of the three-terminal device which is close to the diagram of the metallic loop suggested by R B Laughlin for the explanation of the quantized Hall conductivity (figure 5c) [16].

In conclusions, we have demonstrated experimentally that tuning the Rashba coupling parameter induced by the bias voltage applied to the AB ring with three asymmetrically situated electrodes, it is possible to control efficiently its conductance in the ballistic regime by the measurements of the AC conductance oscillations. The interplay between the Rashba SOI and spontaneous spin polarization of carriers that gives rise to their relative contributions to the spin-dependent transport phenomena in the three-terminal one-dimensional rings have been also established.

We thank M Pepper for useful comments. This work was supported by SNSF in frameworks of the programme "Scientific Cooperation between Eastern Europe and Switzerland, Grant IB7320-110970/1.